\documentclass[10pt,twocolumn,twoside]{IEEEtran}
\usepackage{epsfig}
\usepackage{amssymb}
\usepackage{amsthm}
\usepackage{cite}
\usepackage[cmex10]{amsmath}
\usepackage{color}
\usepackage{epsfig}
\usepackage{url}
\usepackage[T1]{fontenc}

\usepackage{enumerate}
\usepackage{epstopdf}
\usepackage{float}
\usepackage{comment}
\usepackage{enumitem}
\usepackage{textcomp}
\usepackage{etaremune}

\renewcommand{\theenumii}{\roman{enumvi}}

\newtheorem{thm}{Theorem}
\newtheorem{prop}{Proposition}

\newtheorem{cor}{Corollary}

\newcommand*{\suchthat}{\;\ifnum\currentgrouptype=16 \middle\fi|\;}

\allowdisplaybreaks
\renewcommand{\baselinestretch}{0.96}

\begin{document}
\title{Utilization of Water Supply Networks for Harvesting Renewable Energy}

\author{Dariush Fooladivanda, Alejandro D. Dom\'inguez-Garc\'ia, and Peter W. Sauer
\thanks{The authors are with the Department of Electrical and Computer Engineering of the University of Illinois at Urbana-Champaign, Urbana, IL 61801,
USA. Email:\{dfooladi, aledan, psauer\}@ILLINOIS.EDU.}}

\maketitle
\begin{abstract}
\textcolor{black}{Renewable surplus power is increasing due to the increasing penetration of these intermittent resources. In practice, electric grid operators either curtail the surplus energy resulting from renewable-based generations or utilize energy storage resources to absorb it. In this paper, we propose a framework for utilizing water pumps and tanks in water supply networks to absorb the surplus electrical energy resulting from renewable-based electricity generation resources in the electrical grid. We model water supply networks analytically, and propose a two-step procedure that utilizes the water tanks in the water supply network to harvest the surplus energy from an electrical grid. In each step, the water network operator needs to solve an optimization problem that is non-convex. To compute optimal pump schedules and water flows, we develop a second-order cone relaxation and an approximation technique that enable us to transform the proposed problems into mixed-integer second-order cone programs. We then provide the conditions under which the proposed relaxation is exact, and present an algorithm for constructing an exact solution to the original problem from a solution to the relaxed problem. We demonstrate the effectiveness of the proposed framework via numerical simulations.}
\end{abstract}
\IEEEpeerreviewmaketitle

\section{Introduction}

\IEEEPARstart{W}{ater} supply networks are one of the indispensable infrastructures; the availability of drinking water as well as several industrial processes are highly dependent on
the availability of water, as well as the reliability of these networks. Water supply networks consume a substantial amount of electric power to ensure that water demands are delivered at the desired water pressures and qualities. Drinking water and wastewater networks use around four percent of all electricity consumed in the United States \cite{electricity_use}, and water demand is increasing continuously \cite{NG}. To cope with increasing water demands, water network operators aim to minimize their operation costs, and hence they aim at scheduling their water pumps optimally \cite{net_reliability}-\!\cite{pump_group}. Note that water pumps are the major electrical energy consumers in water supply networks.

The optimal pump scheduling and water flow problem can be formulated as a mixed-integer nonlinear program. In this program, energy conservation constraints are non-convex, and some of the decision variables, such as switching pumps on and off, are binary. It is hard to solve this problem over several time periods with reasonable computational time. Due to the computational complexity of this problem, several pump scheduling schemes have been investigated in the literature \cite{Brion}-\!\!\cite{Ghaddar}. 
In \cite{dfooladi1}, the authors focus on the pump scheduling and water flow optimization problem, and propose the first convex relaxation for the feasibility set of this problem. The authors then reformulate the optimal pump scheduling and water flow problem as a mixed-integer second-order cone program that can be solved with commercially-available solvers. They prove that the proposed relaxation is exact for several water network topologies. The proposed convex relaxation and reformulation techniques form a tractable basis for further studies of the interaction between water and energy in water supply networks. \textcolor{black}{In \cite{Josh1}, the authors extend the pump scheduling and water flow problem proposed in \cite{dfooladi1} by formulating a water-power flow problem that optimizes the use of controllable assets across power and water systems. They propose a distributed solver that enables water and power operators to pursue individual objectives while respecting all the physical constraints.}


Wind and solar surplus power is increasing due to the increasing penetration of these intermittent resources \cite{nrel1}. Electric grid operators either curtail the surplus energy resulting from renewable-based generations or utilize energy storage resources to absorb it \cite{nrel2}. In this paper, we focus on the utilization of water pumps and tanks in water supply networks to absorb the surplus electrical energy from the electrical grid, and to harvest the electrical energy in water tanks as potential energy. \textcolor{black}{The benefits of this scheme are twofold: 1) water network operators minimize their electricity costs, and 2) electrical grid operators use renewable energy resources more efficiently.}

\textcolor{black}{The utilization of water supply networks for providing demand response services has been investigated in \cite{dr3}-\!\!\cite{dr5}. In \cite{dr3}, the authors demonstrate how to use the flexibility of water distribution networks for offering energy modulation services. In \cite{dr4}, the authors propose a linear optimization problem for computing the pump schedules. In \cite{dr5}, the authors use a piecewise linear approximation of the head loss formulae given by either the Hazen-William or the Darcy-Weisbach, and propose an optimal pump scheduling problem. Although some approaches in the literature are addressing
the demand response problem in water supply networks, the existing works do approximate the pump characteristic curves and the pump energy costs (e.g., see \cite{dr4} and \cite{dr5}). Hence, the existing frameworks and control schemes lack performance
guarantees in practice.}

We consider a water supply network composed of pipes, junctions, pumps, reservoirs, valves, and tanks, and assume that the water network operator aims to provide demand response services to the grid by scheduling its pumps. More precisely, we focus on harvesting the surplus energy provided by the grid in the water network while ensuring that all the loads are delivered at the desired water pressures. The water network operator also aims at minimizing the operation cost of its water pumps while maintaining water pressures at the desired levels throughout the network. Therefore, the water network operator needs to optimize its pump schedules such that it minimizes its operation cost while utilizing the surplus energy provided by the grid. To achieve this goal, the water network operator needs to participate in energy markets through demand response programs that aim to increase demand at times of high generations and low demand. To participate in a demand response program, the operator needs to compute its offer for day-ahead or real-time energy markets, and submit this offer to the energy market through a demand response program. The computation of the offer for energy markets is not the subject of this paper. Here, we assume that the operator submitted its offer to the market, and it has been cleared to participate; thus, receiving its cleared energy capacity from the market. Note that demand response programs for energy markets operate at time scales of one to ten minutes. For more information on demand response programs, we refer the reader to \cite{dr1}-\!\!\cite{dr2}.

In real-time, the water network operator needs to compute the optimal set-points of its pumps such that the pumps can collectively respond to the demand response signal sent by the independent system operator
(ISO), without violating any of the physical constraints of its network. To compute the optimal set-points, we propose a two-step procedure for harvesting the surplus energy provided by the grid in the water network. In the first step, the water network operator computes the pump schedules and water flows that minimize the operation costs, and then calculates the energy consumption of the pumps using the computed optimal schedule. If the surplus energy is less than or equal to the computed energy consumption, then the water network operator can use the computed schedule and optimize its operation costs; otherwise, the water network operator will need to recompute its pump schedules and water flows in a second step. If the surplus energy is greater than the computed energy consumption, the operator will have access to some surplus energy in addition to its energy consumption. In the second step, the operator computes the pump schedules and water flows that maximize the energy harvested in the tanks ensuring that all the loads are delivered at the desired water pressures. In each of these steps, the operator needs to solve an optimization problem that is non-convex.

Our goal is to optimize the pump schedules in a centralized fashion. To achieve this goal, we need to solve the proposed water flow and pump scheduling optimization problems that are non-convex. In order to obtain tractable optimization problems, we develop a second-order cone relaxation and an approximation technique that enable us to transform the proposed problems into mixed-integer second order cone programs. Such problems can be solved efficiently with commercial solvers such as GUROBI or CPLEX. In addition, several powerful algorithms for solving mixed-integer second-order cone programs exist in the literature \cite{MISCOP1}-\!\!\cite{MISCOP2}. We then provide the conditions under which the proposed relaxation is exact, and present an algorithm for constructing a solution to the original problems from the solutions to the relaxed problem. Finally, we numerically demonstrate the effectiveness of the proposed relaxation and approximation techniques in solving the energy harvesting problem.

\textcolor{black}{Our contributions are as follows. First, we propose a two-step procedure for harvesting the surplus energy provided by the grid in the water network. In each step, the water network operator needs to solve an optimization problem that is non-convex. Second, we develop a second-order cone relaxation and an approximation technique that enable us to transform the proposed problems into tractable programs. We then provide the conditions under which this relaxation is exact, and present an algorithm for constructing a solution to the original problem from a solution to the relaxed problem. Finally, we numerically demonstrate the effectiveness of the proposed solution techniques in solving the energy harvesting problem addressed in this paper.}

\textcolor{black}{The reminder of this paper is organized as follows. The system model is introduced in Section~\ref{sys_sec}. We propose a two-step procedure for harvesting the surplus energy provided by the grid in Section \ref{pb_sec}. Then, we present a second-order cone relaxation and an approximation technique to transform the proposed optimization problems into mixed-integer second-order cone programs in Section \ref{pb_formulation1}. Finally, we numerically demonstrate the effectiveness of our solution techniques in Section \ref{num_sec}. The proofs of all results presented are included in an Appendix.}

\vspace{-0.2cm}

\section{Preliminaries}\label{sys_sec}
\textcolor{black}{In this section, we first introduce the hydraulic model of a water supply network adopted in this work; this model closely follows that used in \cite{dfooladi1}. Next, we present an energy audit of the aforementioned water network. Finally, we describe the demand response market setting and the separation between the time-scales involved in the decision-making for control of water networks.}

\subsection{Water Supply Network Model}\label{h_model}
Consider a water supply network comprised of junctions, tanks, reservoirs, pumps, valves, and pipes, and assume that the water flow through the water supply network is laminar. To differentiate the inlet and outlet of each water tank, we model each tank as a two-node element, one node representing the inlet of the tank and the other node representing the outlet of the tank. \textcolor{black}{More precisely, we introduce fictitious nodes and edges. Fictitious nodes represent the inlets of the tanks, and fictitious edges connect the inlets and outlets of the tanks. Then, the topology of the water network can be represented by a directed graph, $\mathcal{G}(\mathcal{N},\mathcal{E}\cup\mathcal{E}^{(f)})$, where $\mathcal{N}=\mathcal{J}\cup\mathcal{T}\cup\mathcal{T}^{(f)}\cup\mathcal{R}$, with each $i\in \mathcal{J}$ corresponding to a junction, each $i\in \mathcal{T}$ corresponding to the outlet of a tank, each $i\in \mathcal{T}^{(f)}$ corresponding to the inlet of a tank (i.e., $\mathcal{T}^{(f)}$ represents the set of fictitious nodes), and each $i\in\mathcal{R}$ corresponding to a reservoir; and where each entry in the edge set $\mathcal{E}=\{e_1,\cdots,e_{|\mathcal{E}|}\}$ corresponds to a physical element connecting a pair of nodes, and each entry in the edge set $\mathcal{E}^{(f)}=\{e_1^{(f)},\cdots,e_{|\mathcal{E}^{(f)}|}^{(f)}\}$ corresponds to a fictitious edge connecting the inlet and outlet of a tank.} This physical element can potentially be a pipe, a pump, or a valve. The edge set $\mathcal{E}$ can be written as $\mathcal{E}=\mathcal{P}\cup\mathcal{V}\cup\mathcal{L}$ where $\mathcal{P}$, $\mathcal{V}$, and $\mathcal{L}$ denote the sets of pumps, valves, and pipes, respectively. Let $e_a=(i,j)$ be an ordered pair of vertices indicating that edge $e_a=(i,j)$ is incident to vertices $i$ and $j$, and that it carries a non-zero water flow from node $i$ to node $j$. Further, let $\mathcal{N}^{-}_i$ and $\mathcal{N}^{+}_i$ denote the in and out neighbors of node $i$, respectively, in graph $\mathcal{G}$, i.e., $\mathcal{N}^{-}_i=\{j:(j,i)\in\mathcal{E}\}$ $\mathcal{N}^{+}_i=\{j:(i,j)\in\mathcal{E}\}$. 

\subsubsection{Junctions}
Let us assume that time is divided into slots of size $\delta$. Further, let $D_i[k]$ and ${Q_{i,j}[k]}$ (in $\text{m}^3/\text{s}$) denote the water demand at junction $i\in\mathcal{J}$, and the volumetric flow rate through the element $(i,j)\in\mathcal{E}$ at time slot $k$, respectively. We assume that the $D_i[k]$'s are constant over the duration of a time slot, and that at the beginning of each time slot $k$, the water network operator can perfectly estimate $D_i[k]$ for all $i\in\mathcal{J}$. At each junction $i\in\mathcal{J}$, the flow conservation constraint must hold at each time slot $k$:
\begin{align}\small
\label{nodel_flow}\sum_{m\in\mathcal{N}^{-}_i}{Q_{m,i}[k]}-\sum_{j\in\mathcal{N}^{+}_i}Q_{i,j}[k]=D_i[k].
\end{align}
We define $Q[k]=[\{Q_{i,j}[k]\}_{(i,j)\in\mathcal{E}}]^\top\in\mathbb{R}^{|\mathcal{E}|}$ and $D[k]=[\{D_i[k]\}_{i\in\mathcal{J}}]^\top\in\mathbb{R}^{|\mathcal{J}|}$ for all $k$.

Water pressure levels at different nodes enable water movement through the network. Typically, water pressure is assumed proportional to the elevation above a fixed reference point since water is assumed to be incompressible. For each node $i\in\mathcal{N}$, let $H^0_i$ and $H_{i}[k]$ respectively denote the elevation head of node $i$ with respect to a fixed reference point, and the pressure head of node $i$ at time slot $k$. To maintain the pressure head through the network, the network operator has assigned a minimum pressure head $\underline{H}_i$ to each junction $i\in\mathcal{J}$. At each junction $i\in\mathcal{J}$, pressure head $H_{i}[k]$ should satisfy the minimum pressure head constraint at each time slot $k$, i.e.,
\begin{align}\small
\label{node_1}& \underline{H}_i \le H_{i}[k]~.
\end{align}
Here, we define $H[k]=[\{H_j[k]\}_{j\in\mathcal{N}}]^\top\in\mathbb{R}^{|\mathcal{N}|}$.

\subsubsection{Tanks} We model each water tank as a two-node element. Without loss of generality, we assume that the inlet of each water tank is located at the top of the tank. Let $V_i[k]$ denote the volume of water in tank $i\in\mathcal{T}$ at time slot $k$. Then, for each tank $i\in\mathcal{T}$, the water tank balance constraint must hold at all time slots $k$:
\begin{align}\small
\label{tank_1} V_{i}[k]&=V_{i}[k-1]+\delta\sum_{m\in\mathcal{N}^{-}_i}Q_{m,i}[k]-\delta\sum_{j\in\mathcal{N}^{+}_i}{Q_{i,j}[k]},\\
\label{tank_2}0 &\le V_{i}[k] \le \overline{V}_i,
\end{align}
where $V_{i}[0]$ and $\overline{V}_i$ denote the initial volume of water in tank $i$ and the capacity of tank $i$, respectively. We further define $V[k]=[\{V_i[k]\}_{i\in\mathcal{T}}]^\top\in\mathbb{R}^{|\mathcal{T}|}$ for all $k$.

At each time slot $k$, the pressure head change at the outlet of water tank $i\in\mathcal{T}$ can be approximated by\footnote{\textcolor{black}{In this paper, we ignore the kinetic energy and atmospheric pressure through the system as in most water supply network analyses (see, e.g., \cite{energy_audit}).}}
\begin{align}\small
\label{tank_3}& H_{i}[k]-H_{i}[k-1]=\frac{\delta}{A_i}\left(\sum_{m\in\mathcal{N}^{-}_i}Q_{m,i}[k]-\sum_{j\in\mathcal{N}^{+}_i}{Q_{i,j}[k]}\right),
\end{align}
where $H_i[0]$ and $A_i$ denote the initial pressure head of water tank $i$ and the cross-sectional wetted area of tank $i$, respectively. From (\ref{tank_1}) and (\ref{tank_3}), we obtain that $H_i[0]=V_i[0]/A_i$. Since the tank inlet is located at the top of the tank, the pressure head at node $i\in\mathcal{T}^{(f)}$ must be greater than or equal to the elevation head at node $i$, i.e.,
\begin{align}
\label{tank_4}& H_{i}[k] \ge H^0_i.
\end{align}
\textcolor{black}{Typically, the tank inlet is located at the top of the tank, and hence the elevation head at the tank inlet is higher than the elevation head at the tank outlet. To differentiate the inlet and outlet of each water tank, we have defined fictitious nodes in this paper.} For a more detailed discussion on tank models, we refer the reader to \cite{tank_6}.

\subsubsection{Reservoirs} We consider reservoirs as infinite sources of water, and assume that the pressure head on the surface of at each reservoir $i\in\mathcal{R}$ is zero and the water velocity is also zero; therefore, the total head at the reservoir $i$ is equal to the elevation head at node $i$.

\subsubsection{Pumps} The water network operator uses variable speed pumps to pump water through its network. When pump $(i,j)\in\mathcal{P}$ is on, it increases the pressure head between its inlet and outlet; otherwise, it does not allow water to flow through the pump. Let $H_{i,j}[k]$ denote the pressure head gain of the pump $(i,j)$ at time slot $k$; then, $H_{i,j}[k]$ can be computed as follows (e.g., see \cite{pump_group}):
\begin{align}
&\label{H_G} H_{i,j}[k]=a_{i,j} Q_{i,j}^2[k]+b_{i,j} {{Q_{i,j}[k]} \omega_{i,j}[k]}+c_{i,j} \omega_{i,j}^2[k],
\end{align}
where $\omega_{i,j}[k]$ denotes the normalized speed of the pump with respect to its nominal speed. The coefficients $a_{i,j}<0$, $b_{i,j}>0$, and $c_{i,j}>0$ are the pump parameters evaluated at the nominal speed. We define ${\omega}[k]=[\{\omega_{i,j}[k]\}_{(i,j)\in\mathcal{P}}]^\top\in\mathbb{R}^{|\mathcal{P}|}$ for all $k$.

Let $\eta_{i,j}$ denote the electrical-to-hydraulic energy conversion efficiency of pump $(i,j)\in\mathcal{P}$. To obtain $\eta_{i,j}$, we need to compute the motor and pump efficiencies. The motor efficiency depends on the control scheme, rotational speed, and load. In \cite{pump1}, the authors show that commercially available variable frequency drives can maintain high motor efficiencies
over practical ranges of loads and frequencies. They show that for several types of variable frequency drives, the motor efficiency is higher than 97\% at full loads, and that at lower loads, the motor efficiency does not fall below 95\%. Similar results were reported in \cite{pump2}. Based on these observations, in the reminder, we assume motor efficiencies to be constant.

Centrifugal pumps are typically used in variable speed pump stations. For centrifugal pumps, the affinity laws that relate flow, pressure head gain, and power to the speed of the pump, can provide a good approximation of the real pump behavior for a wide range of speeds. However, these laws cannot model the power and efficiency relationships accurately, especially for smaller pumps \cite{pump_eff01}-\!\!\cite{pump_eff00}. Several approximation methods have been investigated in the literature to model the impact of factors that do not scale with the pump speed and depend on the machine size (e.g., see \cite{pump_eff01}, \cite{pump_eff00}). The impact of these parameters on the pump efficiency can be neglected if the changes in speed do not exceed the 33\% of the nominal pump speed \cite{pump_eff00}. This approximation is justified for large pumps  \cite{pump_eff0}-\!\cite{pump_eff00}. Based on this observation, we enforce pump $(i,j)\in\mathcal{P}$ to operate within a certain range of speeds as follows:
\begin{align}
&\label{H_G_speed} \underline{\omega}_{i,j}\le \omega_{i,j}[k] \le \overline{\omega}_{i,j},
\end{align}
where $\overline{\omega}_{i,j}$ and $\underline{\omega}_{i,j}$ denote, respectively, the maximum and minimum allowable speeds of pump $(i,j)$.

Let $y_{i,j}[k]$ be a binary variable that takes value one if the pump $(i,j)$ is on, and takes value zero, otherwise. To maintain the pump efficiency constant, we further enforce pump $(i,j)\in\mathcal{P}$ to operate within a certain range of water flows as follows:
\begin{align}
&\label{pipe_pump2} \underline{Q}_{i,j} y_{i,j}[k]\le Q_{i,j}[k] \le \overline{Q}_{i,j} y_{i,j}[k],
\end{align}
where $\underline{Q}_{i,j}$ and $\overline{Q}_{i,j}$ denote the minimum and maximum allowable water flow through pump $(i,j)\in\mathcal{P}$, respectively. Under the constraints in (\ref{H_G_speed}) and (\ref{pipe_pump2}), the pump efficiency can be assumed to be constant, and hence $\eta_{i,j}$ can be assumed to be constant. Note that the values of $\overline{\omega}_{i,j}$, $\underline{\omega}_{i,j}$, $\underline{Q}_{i,j}$, and $\overline{Q}_{i,j}$ highly depend on the pump characteristics. The operator can calculate the values of these parameters beforehand.

Energy must be conserved between the nodes $i$ and $j$, where $(i,j)\in\mathcal{P}$, when the pump $(i,j)$ carries a non-zero water flow; otherwise, the pressure heads at junctions $i$ and $j$ are decoupled. Thus, the energy conservation constraint across pump $(i,j)\in\mathcal{P}$ can be represented by
\begin{align}
-M_1 (1-y_{i,j}[k])\le(H_{j}[k]+H^0_j)-(H_{i}[k]+H^0_i)-&H_{i,j}[k] \nonumber\\
\label{pipe_pump1}\le M_1 (1-y_{i,j}[k]),&
\end{align}
where $M_1$ and $m_1$ are positive constants; $M_1$ is sufficiently large, and $m_1$ is sufficiently small. Let $E_{i,j}[k]$ denote the energy consumption of pump $(i,j)\in\mathcal{P}$ at time slot $k$, which can be computed as follows:
\[E_{i,j}[k]=\frac{\rho g \delta}{\eta_{i,j}} H_{i,j}[k] Q_{i,j}[k],\]
\textcolor{black}{where $\rho$ and $g$ are the water density and gravity constant (in m/s$^2$), respectively.} Let $\lambda[k]$ (in \$/Wh) denote the electricity price at time slot $k$. Then, the total operation cost at time slot $k$ equals $\sum_{(i,j)\in\mathcal{P}}{\lambda[k] E_{i,j}[k]}$.

\subsubsection{Pressure Reducing Valves}
To control pressure heads at different junctions through the network, the water network operator uses pressure reducing valves (PRV)\footnote{\textcolor{black}{We will use the term valve and PRV interchangeably
in this paper.}}, the actuation of which controls the pressure head losses across the valves. Let $H_{i,j}[k]$ denote the pressure head loss across the valve $(i,j)\in\mathcal{V}$ at time slot $k$; $H_{i,j}[k]$ allows the operator to adjust the head loss from junction $i$ to $j$ as needed. When the valve carries a non-zero water flow, the pressure heads across the valve will be coupled by the energy conservation law; otherwise, the pressure heads across the valve will be decoupled. Let $y_{i,j}[k]$ be a binary variable that takes value one when valve $(i,j)$ carries a non-zero water flow, and zero, otherwise. At each time slot $k$, the energy conservation constraint can be represented by
\begin{align}
-M_1 (1-y_{i,j}[k]) \le (H_{j}[k]+H^0_j)-(H_{i}[k]+H^0_i)&+H_{i,j}[k] \nonumber\\
\label{pipe_valve1}\le M_1 (1-y_{i,j}[k])&,\\
\label{pipe_valve2} m_1 y_{i,j}[k]\le Q_{i,j}[k] \le M_1 y_{i,j}[k].\quad\quad&
\end{align}
Note that the head loss $H_{i,j}[k]$ is a non-negative variable. For more information on valves, we refer the reader to \cite{valv1}-\!\!\cite{valv4}.

\subsubsection{Pipes}
When a pipe $(i,j)\in\mathcal{L}$ carries a non-zero water flow, energy must be conserved between the two junctions $i$ and $j$; otherwise, the pressure heads at junctions $i$ and $j$ are decoupled. The energy conservation constraint across pipe $(i,j)\in\mathcal{L}$ can be represented by
\begin{align}
-M_1 (1-y_{i,j}[k]) \le (H_{j}[k]+H^0_j)-(H_{i}[k]+H^0_i)&+H_{i,j}[k] \nonumber\\
\label{pipe_nopump1}\le M_1 (1-y_{i,j}[k]&),\\
\label{pipe_nopump2} m_1 y_{i,j}[k]\le Q_{i,j}[k] \le M_1 y_{i,j}[k],\quad\quad&
\end{align}
where $y_{i,j}[k]$ is a binary variable that is one if the pipe $(i,j)$ carries a non-zero water flow, and is zero, otherwise. $H_{i,j}[k]$ denotes the pressure head loss of pipe $(i,j)\in\mathcal{L}$ at time slot $k$. We define the vectors $y[k]=[\{y_{i,j}[k]\}_{(i,j)\in\mathcal{E}}]^\top\in\mathbb{R}^{|\mathcal{E}|}$ and $G[k]=[\{H_{i,j}[k]\}_{(i,j)\in\mathcal{V}\cup\mathcal{P}\cup\mathcal{L}}]^\top\in\mathbb{R}^{|\mathcal{V}\cup\mathcal{P}\cup\mathcal{L}|}$ for all $k$.

The Darcy-Weisbach and Hazen-Williams equations are two formulas widely used by researchers to model pressure head losses across pipes (e.g., see \cite{example_1}). Using the Darcy-Weisbach equation, $H_{i,j}[k]$ can be modeled as follows:
\begin{align}
\label{head_loss}H_{i,j}[k]=f^d_{i,j}  {Q_{i,j}^2[k]},
\end{align}
with $f^d_{i,j}={({r_{i,j} \ell_{i,j}})/({2 d_{i,j} s_{i,j}^2 g})}$, where $\ell_{i,j}$ and $d_{i,j}$ are the length and the hydraulic diameter of pipe $(i,j)$ (in meters), respectively, and $r_{i,j}$ and $s_{i,j}$ are the Darcy friction factor and the cross-sectional wetted area of pipe $(i,j)$, respectively. Using the Hazen-Williams equation, $H_{i,j}[k]=f^h_{i,j}  {Q_{i,j}^{1.852}[k]}$ with $f^h_{i,j}={({r_{i,j} \ell_{i,j}})/({c_{i,j}^{4.871} d_{i,j}^{4.871}})}$ where $r_{i,j}$ and $c_{i,j}$ are the head loss coefficient factor and the roughness of pipe $(i,j)$, respectively. In this paper, we use the Darcy-Weisbach equation to compute the $H_{i,j}[k]$'s.


\subsection{Energy Audit of Water Networks}\label{energyAudit}
We now present an energy audit of the water network for one single time slot of length $\delta$. To do so, we select the water network except its pumps and reservoirs as a control volume, and consider the set of pumps and reservoirs as the boundaries of our control volume. These boundaries determine which elements are externally contributing to the energy flow or internally storing or dissipating energy. In addition, we make the following assumptions:
\begin{enumerate}
\item[A.1] The heat flow through the boundaries is zero.
\item[A.2] The mechanical work is supplied by the pumps.
\end{enumerate}
Under the assumptions above, we can calculate the energy contribution of different water network elements to the control volume as follows:
\begin{itemize}
\item The external energy provided by the pumps is equal to $E_p[k]={\rho g \delta}{ \sum_{(i,j)\in\mathcal{P}}{ { H_{i,j}[k] Q_{i,j}[k]} } }$.
\item The external energy supplied by the reservoirs is equal to $E_r[k]=\rho g \delta\sum_{i\in\mathcal{R}}\sum_{j\in\mathcal{N}^{+}_i}Q_{i,j}[k] H^0_i$.
\item The internal energy dissipated to overcome the elevation head differences at different nodes and the friction head losses of the pipes and PRVs equals $E_l[k]=\rho g \delta\sum_{(i,j)\in\mathcal{E}\setminus\mathcal{P}}
{Q_{i,j}[k]({H}_j[k]-{H}_i[k])}$.
\item The energy supplied or absorbed by the tanks equals  $E_t[k]=\rho g \delta \sum_{i\in\mathcal{T}} (A_i/2) \left(H_{i}^2[k]-H_{i}^2[k-1]\right)$.
\item The energy delivered to the consumers is equal to $E_d[k]=\rho g \delta\sum_{i\in\mathcal{J}}{ {{{D_{i}[k]}} H_{i}[k]}}$.
\end{itemize}

We apply the energy conservation law to the control volume with known amounts of water and energy flowing through the boundaries. At each time slot $k$, the energy conservation constraint can be represented by
\begin{align}
\label{energy_balance}&E_p[k]+E_r[k]=E_t[k]+E_l[k]+E_d[k];\nonumber
\end{align}
i.e., the energy supplied to the water network through pumps and reservoirs equals the sum of (i) the energy needed to overcome the elevation head differences at different nodes and the friction head losses of the pipes, (ii) the energy supplied or absorbed by the tanks, and (iii) the energy delivered to the consumers. For more information on the energy audit of water networks, we refer the reader to \cite{energy_audit}.


\subsection{Demand Response Market Mechanism}
The water network operator participates in an energy market through a demand response program that is designed to balance the power generation and load by changing electricity demands. Demand response programs require their resources to submit their offers to energy markets before the contract period starts \cite{dr1}-\!\!\cite{dr2}. Therefore, the water network operator needs to compute its offer for day-ahead or real-time energy markets, and to submit the offer to the energy market through the demand response program. Note that the operator offer for the energy market will consist of the maximum power that can be absorbed by the water pumps at each time slot $k$. The duration of the time slot for demand response programs in energy markets ranges from one to ten minutes \cite{dr1}-\!\!\cite{dr2}. At such time scales, water pumps can follow the demand response signal sent by the ISO.

In this paper, we assume that the water network operator computed its demand response capacity for day-ahead or real-time energy markets, and submitted the offer to the energy market through a demand response program. We further assume that the water network has been selected to provide a certain amount of demand response capacity. We denote this quantity by $\overline{r}$. We assume that the water network operator's offer is computed beforehand. Hence, the water network operator's capacity $\overline{r}$ is fixed and known.
\textcolor{black}{The computation of the water network operator's offer for day-ahead or real-time energy markets is not the subject of this work.}

Let us assume that the contract duration is $T=K \delta$. During the contract, the ISO sends a new demand response signal to the water network operator at every time slot $k\in\mathcal{K}$ where $\mathcal{K}=\{1,\cdots,K\}$. The water network operator and ISO has agreed on the capacity $\overline{r}$ (in W) for the contract duration $T$. The ISO has committed to send some demand response signal $r[k]$ at
time slot $k$ that satisfy the following constraint:
\[
0\le r[k]\le \overline{r}~.
\]
The water network operator has committed to draw constant power $r[k]$ at time slot $k$ from the grid.

\textcolor{black}{The water network operator will be rewarded for the amount of energy that it draws during the contract, and hence it will be penalized for any failure in responding to the demand response signals $r[1],\cdots,r[K]$. \textcolor{black}{The water network operator can maximize its net revenue by minimizing the cost of failing to follow some demand response signals over the contract period $T$. Without loss of generality, we take the cost of failing to follow some demand response signal to be linearly proportional to the imbalance $|\gamma[k]-r[k]|$, where $\gamma[k]$ denotes the total power that is drawn from the grid by the water network operator at time slot $k$. Hence, the cost of failing to follow the demand response signal over the contract duration $T$ equals $\sum_{k=1}^K{\lambda[k]|\gamma[k]-r[k]|}$ (\text{in \$}), where $\lambda[k]$ denotes the price of electricity at time slot $k$. Notice that $\lambda[k]$ is a stochastic price parameter ($\$/\text{W}$) modeling the price of electricity in the real-time market.}}

\subsection{Time-scale Separation Principle}
The water network operator controls its resources at different time-scales \cite{struc1}-\!\!\cite{struc2}. First, at a slower time-scale (e.g., day-ahead), the operator computes the optimal set-points of its pumps, PRVs, and tanks (especially the values of binary variables $y_{i,j}[k]$'s) given a set of historic loads and electricity prices. These set-points will determine the water network's topology for the next 24 hours, i.e., the computed set-points will determine which pipes and pumps will carry non-zero water flows during the next 24 hours. Then, at a faster time-scale (e.g., real-time), the operator computes the optimal schedules of its tanks, pumps, and valves at each time slot $k$ given the values of water demand, $D[k]$, electricity price, $\lambda[k]$, and control variable, $y[k]$.

In this paper, we focus on a single demand response contract in the energy market, and assume that the values of the binary variables, i.e., the $y_{i,j}[k]$'s, are fixed and known to the water network operator. \textcolor{black}{We further assume that the value of each binary variable $y_{i,j}[k]$ does not change over the contract period $T$, i.e., for each $(i,j)\in\mathcal{E}$, $y_{i,j}[k]$ is either zero or one for all $k\in\mathcal{K}$.} Given the $y_{i,j}[k]$'s, we can now construct a subgraph of the directed graph \textcolor{black}{$\mathcal{G}(\mathcal{N},\mathcal{E}\cup\mathcal{E}^{(f)})$ as follows. First, we remove all the edges $e_a=(i,j)$ with $y_{i,j}[k]=0$ from $\mathcal{G}(\mathcal{N},\mathcal{E}\cup\mathcal{E}^{(f)})$, and then remove all the isolated nodes from the resulting graph. We call the new graph $\mathcal{G}_y(\mathcal{N}_y,\mathcal{E}_y\cup\mathcal{E}^{(f)}_y)$, where $\mathcal{N}_y\subset\mathcal{N}$, $\mathcal{E}_y\subset\mathcal{E}$, and $\mathcal{E}^{(f)}_y\subset\mathcal{E}^{(f)}$. {The graph $\mathcal{G}_y$ is fixed over the contract period $T$, and determines the set of water network elements that are online, as well as their connectivity.} Let $\mathcal{M}$ and $\mathcal{U}_i$ denote the set of junctions with multiple incoming pipes and the set of upstream nodes of node $i\in\mathcal{N}_y$, respectively, in graph $\mathcal{G}_y$. Further, let $\mathcal{N}_{i,j}$ denote the set of tanks, and junctions with multiple incoming pipes over all the paths between nodes $i$ and $j$ including $j$ in graph $\mathcal{G}_y$. Next, we formulate the problem of controlling the pumps, PRVs, and tanks in the water network at every time slot $k$.} 


\textcolor{black}{We now focus on an online setting in which at the beginning of each time slot $k$, the value of water demand $D[k]$, and electricity price $\lambda[k]$ are fixed and known to the water network operator. At the beginning of each time slot $k$, the operator receives the value of a demand response signal, $r[k]$, from the ISO, and schedules its pumps and tanks so that they can collectively harvest the surplus energy provided by the power grid while minimizing the operation cost of the pumps at time slot $k$. Next, we formulate the problem of controlling the pumps, PRVs, and tanks in the water network at every time slot $k$.}

\section{Problem Statement}\label{pb_sec}
Consider a single demand response contract in the energy market. The water network operator aims at harvesting the surplus energy provided by the power grid in its water network while ensuring that all the loads are delivered at the desired pressure heads. In addition, the operator aims at minimizing the operation cost of its water pumps while maintaining the pressure heads at the desired levels through the network. Therefore, the operator needs to optimize its pump schedules such that it minimizes its operational cost while utilizing the surplus energy provided by the power grid. To achieve this goal, we focus on an online setting in which at the beginning of each time slot $k$, the values of $D[k]$, $\lambda[k]$, and $r[k]$ are revealed to the water network operator, and propose a two-step procedure for computing the optimal schedules of pumps, tanks, and valves as follows:

\textbf{Step 1:} The water network operator computes its minimum operation costs at time slot $k$ by computing a solution to the following problem. Let us define the vector $x[k]=[H[k]^\top,Q[k]^\top,G[k]^\top,V[k]^\top]^\top$. Given the parameters $\overline{V}_i$, $\underline{H}_i$, $H^0_i$, $\overline{\omega}_{i,j}$, $\eta_{i,j}$, $f^d_{i,j}$, and $D[k]$, the computation of $x[k]$ and $\omega[k]$ can be accomplished by solving the following optimization problem:
\begin{align}
&{\textbf{M}}_{1}:\quad\min_{{x[k]},{\omega[k]}}{\sum_{(i,j)\in\mathcal{P}}{\lambda[k] \frac{\rho g \delta}{\eta_{i,j}} H_{i,j}[k] Q_{i,j}[k]}}\nonumber\\
&\quad\text{subject}~\text{to}~(\ref{nodel_flow})-(\ref{head_loss}).\nonumber
\end{align}
Note that the coefficient $\rho g \delta\lambda[k]$ can be removed from the objective function since it is fixed and known.

Let $x^*[k]$ and $\omega^*[k]$ denote the optimal solution to ${\textbf{M}}_{1}$, and let $E_p^*[k]$ denote the energy consumption of the pumps using the schedules $x^*[k]$ and $\omega^*[k]$. If $E_p^*[k]\ge r[k] \delta$, then the water network operator needs to purchase the energy imbalance $(E_p^*[k]-r[k]\delta)$ from the electricity market to ensure that all the loads are delivered at the desired pressure heads. In this case, the water network operator can minimize its operation costs while following the demand response signal, by using the schedules $x^*[k]$ and $\omega^*[k]$. However, if $E_p^*[k]< r[k]\delta$, the schedules $x^*[k]$ and $\omega^*[k]$ may not efficiently harvest the surplus energy $(r[k]\delta-E_p^*[k])$ in the tanks since our objective in ${\textbf{M}}_{1}$ is to minimize the operation cost at time slot $k$. As mentioned in Section \ref{energyAudit}, the energy supplied to the water network through pumps and reservoirs equals the sum of the energy needed to overcome the elevation head differences at different nodes and the  head losses of the pipes, the energy supplied or absorbed by the tanks, and the energy delivered to the consumers. Therefore, the schedules $x^*[k]$ and $\omega^*[k]$ do not necessarily maximize the energy harvested in the tanks at time slot $k$. Hence, if $E_p^*[k]< r[k]\delta$, the operator will need to compute the schedules $x[k]$ and $\omega[k]$ that maximize the energy absorbed by the tanks at time slot $k$, in the second step.

\textbf{Step 2:} Let us assume that the surplus energy is greater than the optimal energy consumption, i.e., $E_p^*[k]< r[k]\delta$. To maximize the energy harvested in the tanks, we select the energy supplied or absorbed by the tanks as our objective function, and maximize the energy harvested at time slot $k$ by computing a solution as follows. Recall that the water tanks are the elements that can absorb the surplus energy provided by the grid. Given the parameters $\overline{V}_i$, $\underline{H}_i$, $H^0_i$, $\overline{\omega}_{i,j}$, $\eta_{i,j}$, $f^d_{i,j}$, and $D[k]$, the computation of $x[k]$ and $\omega[k]$ can be obtained by solving the following optimization problem:
\begin{align}
{\textbf{N}}_{1}:\quad&\max_{{x[k]},{\omega[k]}}{ \rho g \delta \sum_{i\in\mathcal{T}} \frac{A_i}{2} \left(H_{i}^2[k]-H_{i}^2[k-1]\right)}\nonumber\\
&\quad\text{subject}~\text{to}~(\ref{nodel_flow})-(\ref{head_loss})\nonumber\\
\label{sum_bilin}&\sum_{(i,j)\in\mathcal{P}}{\frac{\rho g \delta}{\eta_{i,j}} H_{i,j}[k] Q_{i,j}[k]}\le r[k]\delta,
\end{align}
where the $H_{i}[k-1]$'s are given. The constraint in (\ref{sum_bilin}) ensures that the energy consumption of the pumps will be less than or equal to $r[k]\delta$.

Problems ${\textbf{M}}_{1}$ and ${\textbf{N}}_{1}$ are non-convex due to the non-convex objective functions and feasibility regions. Hence, it is hard to solve these problems exactly in their current forms. Our goal is to compute exact solutions to ${\textbf{M}}_{1}$ and ${\textbf{N}}_{1}$. To achieve this goal, we focus on the feasibility region and objective function of the energy harvesting maximization problem ${\textbf{N}}_{1}$, and transform ${\textbf{N}}_{1}$ into a mixed-integer second-order cone program that can be solved efficiently with commercial solvers. In addition, we propose a set of sufficient conditions under which the proposed relaxation is exact. The proposed relaxation and reformulation techniques can be applied to the operation cost minimization problem ${\textbf{M}}_{1}$ since problems ${\textbf{M}}_{1}$ and ${\textbf{N}}_{1}$ have similar structures. In what follows, we only present the results for the energy harvesting maximization problem ${\textbf{N}}_{1}$.

\section{Energy Harvesting Maximization Problem: Reformulation and Relaxation}\label{pb_formulation1}
The energy harvesting problem ${\textbf{N}}_{1}$ is non-convex because its feasible set is non-convex. We follow a three-step procedure to transform problem ${\textbf{N}}_{1}$ into a mixed-integer second-order cone program that can be solved with commercial solvers. To do so, we start with the pump hydraulic constraints, and convexify the feasible set associated with pump variables. Using the proposed convex relaxation, we formulate a new optimal pump scheduling and water flow problem referred to as ${\textbf{N}}_{2}$. Second, we focus on the nonlinear equality constraint (\ref{head_loss}), and propose a convex relaxation for constraint (\ref{head_loss}) referred to as ${\textbf{N}}_{3}$. Finally, we focus on the non-convex quadratic constraint in (\ref{sum_bilin}), and propose an inner approximation for this constraint. In addition, we approximate the quadratic terms $H_i^2[k]$'s for all $i\in\mathcal{T}$, and transform the non-convex program ${\textbf{N}}_{3}$ into a mixed-integer second-order cone program referred to as ${\textbf{N}}_{4}$. These steps enable us to transform problem ${\textbf{N}}_{1}$ into a mixed-integer second-order cone program. The solution to ${\textbf{N}}_{4}$ provides an approximation to an exact solution to ${\textbf{N}}_{1}$.


\subsection{Convexification of the Pump Hydraulic Constraints}
\begin{figure}[t]
\begin{center}
\includegraphics[width=2.8in]{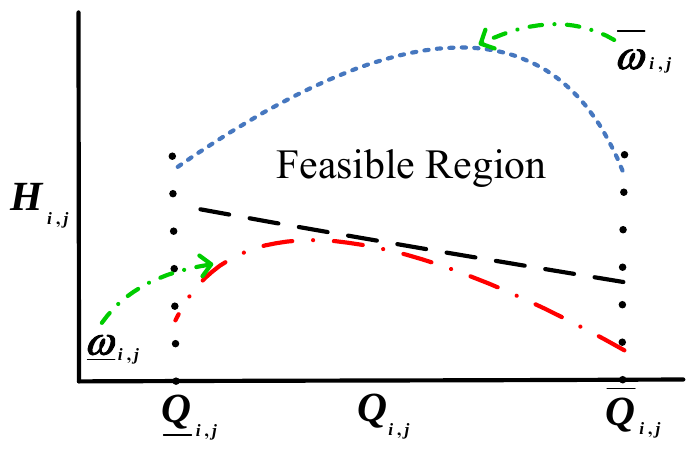} \caption{A graphical depiction of pump $(i,j)\in\mathcal{P}$ feasibility region: The dotted line and the dash-dot line reprsenet the set of feasible values of pair $(Q_{i,j}[k],H_{i,j}[k])$ for ${\omega}_{i,j}[k]=\overline{\omega}_{i,j}$ and ${\omega}_{i,j}[k]=\underline{\omega}_{i,j}$, respectively. The dashed line represents the line ${d}_{i,j} {{Q_{i,j}[k]}}+{e}_{i,j}$. The region between the dashed line and the dotted line above represents the convex feasible set determined by (\ref{new_H_G11}) and (\ref{new_H_G1}) while the area between the dotted line and the dash-dot line represents the original feasibility set of the pump constraints.} \label{sconvexFig}
\end{center}
\end{figure}

We focus on the pump hydraulic constraints in (\ref{H_G})-(\ref{pipe_pump2}), and represent the feasibility set of the pump constraints in the space of variables $Q_{i,j}[k]$ and $H_{i,j}[k]$. Note that we can reduce the dimension of the pump feasibility set from $(\omega_{i,j}[k],Q_{i,j}[k],H_{i,j}[k])$ to $(Q_{i,j}[k],H_{i,j}[k])$ due to the structure of constraints (\ref{H_G}) and (\ref{H_G_speed}). The feasibility set of the pump constraints is non-convex, as shown in Fig. \ref{sconvexFig}. In particular, the boundary of the feasibility set obtained by the equation $H_{i,j}[k]= a_{i,j} Q_{i,j}^2[k]+{b}_{i,j} \underline{\omega}_{i,j} {{Q_{i,j}[k]}}+\underline{\omega}_{i,j}^2{c}_{i,j}$ makes the feasibility set non-convex. We approximate this constraint with a linear constraint, and propose an inner convex approximation for the pump hydraulic constraints in (\ref{H_G})-(\ref{pipe_pump2}) as follows:
\begin{align}
\label{new_H_G11}&H_{i,j}[k]\le a_{i,j} Q_{i,j}^2[k]+\overline{b}_{i,j} {{Q_{i,j}[k]}}+\overline{c}_{i,j},~\forall (i,j)\in\mathcal{P},\\
\label{new_H_G1}&H_{i,j}[k]\ge {d}_{i,j} {{Q_{i,j}[k]}}+{e}_{i,j},~\forall (i,j)\in\mathcal{P},
\end{align}
where $\overline{b}_{i,j}=b_{i,j} \overline{\omega}_{i,j}$ and $\overline{c}_{i,j}=c_{i,j} {{\overline{\omega}^2_{i,j}}}$. ${d}_{i,j}$ and ${e}_{i,j}$ are chosen so that the following inequality holds for all values of $Q_{i,j}[k]$:
\begin{align}
{d}_{i,j} {{Q_{i,j}[k]}}+{e}_{i,j} \ge a_{i,j} Q_{i,j}^2[k]+{b}_{i,j} \underline{\omega}_{i,j} {{Q_{i,j}[k]}}+\underline{\omega}_{i,j}^2{c}_{i,j}~. \nonumber
\end{align}
The constraint (\ref{new_H_G11}) is convex since $a_{i,j}$ is negative. The dashed line in Fig.~\ref{sconvexFig} represents the line ${d}_{i,j} {{Q_{i,j}[k]}}+{e}_{i,j}$; the values of the parameters ${d}_{i,j}$ and ${e}_{i,j}$ are highly dependent on the real pump behavior, and can be computed beforehand.

We now formulate a new pump scheduling and water flow problem as follows: Given the system parameters, $x[k]$ can be obtained by solving the following optimization problem:
\begin{align}
&{\textbf{N}}_{2}:\quad\quad\max_{x[k]}~{\sum_{i\in\mathcal{T}} A_i H_{i}^2[k]}\nonumber\\
&\quad\quad\text{subject}~\text{to}~(\ref{nodel_flow})-(\ref{tank_4}), (\ref{pipe_pump2})-(\ref{head_loss}), (\ref{sum_bilin}), (\ref{new_H_G11})-(\ref{new_H_G1})\nonumber
\end{align}
Note that the coefficient $\rho g \delta$ and the quadratic terms $H_{i}[k-1]$'s are removed from the objective function since they are fixed and known at the beginning of time slot $k$. The solution to problem ${\textbf{N}}_{2}$ provides a lower bound on the optimal value of the objective function in ${\textbf{N}}_{1}$.




We can compute the optimal values of $\omega_{i,j}[k]$'s using the optimal solution to ${\textbf{N}}_{2}$ and the relation in (\ref{H_G}) as follows:
\begin{align}
&\omega_{i,j}^*[k]=\frac{-b_{i,j}{Q_{i,j}^*[k]} {+\atop-} \sqrt{\Delta }}{2 c_{i,j}},\nonumber
\end{align}
where $\Delta=b_{i,j}^2(Q_{i,j}^*[k])^2-4 c_{i,j} (a_{i,j}{(Q_{i,j}^*[k])^2}-H_{i,j}^*[k])$, and $Q_{i,j}^*[k]$ and $H_{i,j}^*[k]$ denote the optimal solution to ${\textbf{N}}_{2}$. Note that given $H_{i,j}[k]$ and $Q_{i,j}[k]$, there exists only one non-negative solution  to the pump hydraulic constraint (\ref{H_G}) \cite{dfooladi1}. In what follows, we work with ${\textbf{N}}_{2}$.

\vspace{-0.2cm}
\subsection{Second-Order Cone Relaxation}\label{relax_thrm_sec}
We now propose a second-order cone relaxation for the quadratic equality constraint (\ref{head_loss}), and formulate a new problem as follows: Given the system parameters, the computation of $x[k]$ can be accomplished by solving the following optimization problem:
\begin{align}
{\textbf{N}}_{3}:&\quad\max_{x[k],w[k]}{\sum_{i\in\mathcal{T}} {A_i} H_{i}^2[k]}\nonumber\\
\quad\text{subject}&~\text{to}~(\ref{nodel_flow})-(\ref{tank_4}), (\ref{pipe_pump2})-(\ref{pipe_nopump2}), (\ref{sum_bilin}), (\ref{new_H_G11})-(\ref{new_H_G1})\nonumber\\
\label{SOCP_1}&H_{i,j}[k]=f^d_{i,j} {{W_{i,j}[k]}},~\forall(i,j)\in\mathcal{L},\\
\label{SOCP_2}& {Q_{i,j}^2[k]} \le W_{i,j}[k],~\forall(i,j)\in\mathcal{L},
\end{align}
where $w[k]=[\{W_{i,j}[k]\}_{(i,j)\in\mathcal{L}}]^\top\in\mathbb{R}^{|\mathcal{L}|}$ for all $k$. The second-order cone relaxation of the friction head losses was originally proposed in \cite{dfooladi1}.

In this paper, we consider two problems \emph{equivalent} if an optimal solution of one can be obtained from an optimal solution of the other. The following result provides a set of sufficient conditions under which either ${\textbf{N}}_{3}$ is exact or an exact solution to ${\textbf{N}}_{2}$ can be constructed from an exact solution to ${\textbf{N}}_{3}$ (i.e., the conditions under which ${\textbf{N}}_{2}$ and ${\textbf{N}}_{3}$ are equivalent). The proof is provided in the appendix.

\begin{thm}\label{thrm3}
\textcolor{black}{Let $\mathcal{G}_y(\mathcal{N}_y,\mathcal{E}_y\cup\mathcal{E}^{(f)}_y)$} denote the graph of a water network. Given the parameters $\overline{V}_i$, $\underline{H}_i$, $H^0_i$, $\overline{\omega}_{i,j}$, $\eta_{i,j}$, $f^d_{i,j}$, and $D[k]$, ${\textbf{N}}_{3}$ is equivalent to ${\textbf{N}}_{2}$ if the following conditions are satisfied:
\begin{enumerate}
\item \textcolor{black}{Graph $\mathcal{G}_y(\mathcal{N}_y,\mathcal{E}_y\cup\mathcal{E}^{(f)}_y)$ is loop-free.}
\item All the incoming pipes in each junction $j\in\mathcal{J}$ with multiple incoming pipes are equipped with PRVs.
\end{enumerate}
\end{thm}

Under these conditions above, either ${\textbf{N}}_{3}$ is exact or an exact solution to ${\textbf{N}}_{2}$ can be constructed from an exact solution to ${\textbf{N}}_{3}$. When ${\textbf{N}}_{3}$ is not exact and the conditions of Theorem \ref{thrm3} are satisfied, the water network operator needs to construct an exact solution to ${\textbf{N}}_{2}$ from a solution to ${\textbf{N}}_{3}$ in order to control the tanks, pumps, and valves optimally; Corollary \ref{cor1} enables us to construct such solutions. The proof of this result closely follows the proof of Theorem \ref{thrm3}. Due to space limitations, we do not include it.

\begin{cor}\label{cor1}
Given the parameters $\overline{V}_i$, $\underline{H}_i$, $H^0_i$, $\overline{\omega}_{i,j}$, $\eta_{i,j}$, $f^d_{i,j}$, and $D[k]$, let $x^*[k]$ and $w^*[k]$ denote the optimal solution to ${\textbf{N}}_{3}$. Consider vectors $x^{(r)}[k]$ and $w^{(r)}[k]$ with the following entries:
\begin{align}\small
V^{(r)}[k]&=V^*[k],~Q^{(r)}[k]=Q^*[k],\nonumber\\
{H_{i}^{(r)}[k]}&={H_{i}^*[k]}+\psi_{i}[k],~\forall i\in\mathcal{N}_y,\nonumber\\
{W_{i,j}^{(r)}[k]}&={({Q^*_{i,j}[k]})}^2,~\forall(i,j)\in\mathcal{L},\nonumber\\
H_{i,j}^{(r)}[k]&=f^d_{i,j} {({Q^*_{i,j}[k]})}^2,~\forall(i,j)\in\mathcal{L},\nonumber\\
H_{i,j}^{(r)}[k]&=H_{i,j}^*[k],~\forall(i,j)\in\mathcal{P},\nonumber\\
{H}^{(r)}_{i,j}[k]&=\left\{ \begin{array}{ll}
{H_{i,j}^*[k]}+\psi_{i}[k],~\text{if}~j\in\mathcal{M} ~\text{and}~ (i,j)\in\mathcal{V}\nonumber\\
{H_{i,j}^*[k]},~\text{if}~j\notin\mathcal{M}~\text{and}~ (i,j)\in\mathcal{V}\nonumber
\end{array} \right.\nonumber
\end{align}
with $\psi_{i}[k]=\sum_{(i',j')\in\mathcal{E}_{i}}{\epsilon_{i',j'}[k]}$ where the scalar $\epsilon_{i,j}[k]$ and the set $\mathcal{E}_{i,j}$ are defined as follows:
\begin{align}
&\epsilon_{i,j}[k]=f^d_{i,j} ({W_{i,j}^*[k]}-{({Q^*_{i,j}[k]})}^2),~ \forall(i,j)\in\mathcal{L},\nonumber\\
&\mathcal{E}_{i}=\{(i',j')|(i',j')\in\mathcal{L},~i\in\mathcal{D}_{i'},~\mathcal{N}_{j',i}=\emptyset\},~\forall i\in\mathcal{N}_y.\nonumber
\end{align}
Then, $x^{(r)}[k]$ and $w^{(r)}[k]$ form an exact solution to ${\textbf{N}}_{3}$.
\end{cor}
The result above allows the water network operator to construct an exact solution to ${\textbf{N}}_{2}$ from a solution to ${\textbf{N}}_{3}$ by updating the pressure head losses across the PRVs. This procedure requires the computation of the set $\mathcal{E}_{i,j}$, defined in Corollary \ref{cor1}, for all $(i,j)\in\mathcal{V}$. As mentioned earlier, we assume that \textcolor{black}{graph $\mathcal{G}_y(\mathcal{N}_y,\mathcal{E}_y\cup\mathcal{E}^{(f)}_y)$} is fixed and known to the water network operator.
Hence, the operator can easily compute the set $\mathcal{E}_{i,j}$ for all $(i,j)\in\mathcal{V}$ in advance.

\vspace{-0.15cm}
\subsection{Discretization and Approximation}
We now focus on the convex quadratic objective function and the non-convex quadratic constraint in (\ref{sum_bilin}), and approximate the quadratic terms $H_{i,j}[k] Q_{i,j}[k]$'s for all $(i,j)\in\mathcal{P}$ and $H_i^2[k]$'s for all $i\in\mathcal{T}$ in order to obtain a tractable optimization problem. To do so, we discretize the $H_{i,j}[k]$'s for all $(i,j)\in\mathcal{P}$, and the $H_i[k]$'s for all $i\in\mathcal{T}$. We further define a new set of auxiliary variables that enable us to represent the non-convex quadratic constraint and the objective function with a set of mixed-integer linear constraints.

Let $\mathcal{Z}=\{\zeta_0,\cdots,\zeta_B\}$ denote the set of values that the $H_{i,j}[k]$, $(i,j)\in\mathcal{P}$, can take, and let $\mathcal{B}=\{1,\cdots,B\}$ be the index set of $\mathcal{Z}$. For each $b\in\mathcal{B}$, let $z_{i,j,b}[k]$ denote a binary variable taking value one if $\zeta_{b-1}\le H_{i,j}[k]<\zeta_{b}$, and zero, otherwise. For each pump $(i,j)\in\mathcal{P}$, we discretize $H_{i,j}[k]$ as follows:
\begin{align}
\label{disct1}&H_{i,j}[k]=\sum_{b\in\mathcal{B}} {z_{i,j,b}[k] \zeta_b},\\
&y_{i,j}[k]=\sum_{b\in\mathcal{B}} {z_{i,j,b}[k]},\\
&z_{i,j,b}[k] \zeta_{b-1} \le H_{i,j}[k],~\forall b\in\mathcal{B},\\
\label{disct4}&H_{i,j}[k] \le z_{i,j,b}[k] \zeta_{b}+(1-z_{i,j,b}[k] )M_1,~\forall b\in\mathcal{B}.
\end{align}
Recall that $M_1$ is sufficiently large, and $y_{i,j}[k]$ is a parameter which is one if the pump $(i,j)$ is on, and is zero, otherwise. Using the discretization technique in (\ref{disct1})-(\ref{disct4}), we obtain:
\begin{align}
&{\sum_{(i,j)\in\mathcal{P}}{ \frac{H_{i,j}[k] Q_{i,j}[k]}{\eta_{i,j}}}}=\sum_{(i,j)\in\mathcal{P}}{\frac{1}{\eta_{i,j}} \sum_{b\in\mathcal{B}} {\Phi_{i,j,b}[k] \zeta_b}},\\
\label{lin0}&\Phi_{i,j,b}[k]=Q_{i,j}[k] z_{i,j,b}[k],~\forall b\in\mathcal{B},
\end{align}
where $\Phi_{i,j,b}[k]$ is an auxiliary variable. The result in the following proposition, established in \cite{glover}, enables us to linearize the quadratic constraint (\ref{lin0}), and represent the non-convex quadratic constraint in (\ref{sum_bilin}) with a set of linear constraints.


\begin{prop}\label{glov}
Let $\mathcal{Y}\subset\mathbb{R}$ be a compact set. Given a binary variable $x$ and a linear function $g(y)$ in a continuous variable $y\in \mathcal{Y}$, $z$ equals the quadratic function $xg(y)$ if and only if
\begin{align}
&\underline{g}x \le z \le \overline{g}x,\nonumber\\
&g(y)-\overline{g}(1-x) \le z\le g(y)-\underline{g}(1-x),\nonumber
\end{align}
where $\underline{g}=\min_{y\in \mathcal{Y}}\{g(y)\}$ and $\overline{g}=\max_{y\in \mathcal{Y}}\{g(y)\}$.
\end{prop}
Using Proposition \ref{glov},
(\ref{lin0}) can be represented by
\begin{align}
&\label{lin1}\underline{Q}_{i,j} z_{i,j,b}[k]\le \Phi_{i,j,b}[k]\le \overline{Q}_{i,j} z_{i,j,b}[k],~\forall b\in\mathcal{B},\\
&Q_{i,j}[k]-(1-z_{i,j,b}[k])\overline{Q}_{i,j} \le \Phi_{i,j,b}[k],~\forall b\in\mathcal{B},\\
&\label{lin4}\Phi_{i,j,b}[k] \le Q_{i,j}[k]-(1-z_{i,j,b}[k])\underline{Q}_{i,j},~\forall b\in\mathcal{B}.
\end{align}
Recall that $\underline{Q}_{i,j}$ and $\overline{Q}_{i,j}$ denote the minimum and maximum allowable water flow through pump $(i,j)\in\mathcal{P}$, respectively. We define $z[k]=[\{z_{i,j,b}[k]\}_{(i,j)\in\mathcal{P},b\in\mathcal{B}}]^\top\in\mathbb{R}^{B|\mathcal{P}|}$ and $\Phi[k]=[\{\Phi_{i,j,b}[k]\}_{(i,j)\in\mathcal{P},b\in\mathcal{B}}]^\top\in\mathbb{R}^{B|\mathcal{P}|}$.

To discretize pressure heads, let $\mathcal{S}=\{\sigma_0,\cdots,\sigma_C\}$ denote the set of values that the pressure head $H_{i}[k]$, where $i\in\mathcal{T}$, can take, and let $\mathcal{C}=\{1,\cdots,C\}$ be the index set of $\mathcal{S}$. For each $c\in\mathcal{C}$, let $s_{i,c}[k]$ be a binary variable which is one if $\sigma_{c-1}\le H_i[k]<\sigma_{c}$, and let it be zero, otherwise. Let us define the vector $s[k]=[\{s_{i,c}[k]\}_{i\in\mathcal{T},c\in\mathcal{C}}]^\top\in\mathbb{R}^{C|\mathcal{T}|}$. For each tank $i\in\mathcal{T}$, we discretize pressure head $H_i[k]$ as follows:
\begin{align}
\label{disct11}&H_i[k]=\sum_{c\in\mathcal{C}} {s_{i,c}[k] \sigma_c},\\
&\sum_{c\in\mathcal{C}} {s_{i,c}[k]}=1,\\
&s_{i,c}[k] \sigma_{c-1} \le H_i[k],~\forall c\in\mathcal{C},\\
\label{disct41}&H_i[k] \le s_{i,c}[k]\sigma_{c}+(1-s_{i,c}[k] )M_1,~\forall c\in\mathcal{C}.
\end{align}

We now reformulate ${\textbf{N}}_{3}$ as follows:
\begin{align}
{\textbf{N}}_{4}:&\quad\max_{x[k],w[k],\rho[k]}{\sum_{i\in\mathcal{T}} A_i \sum_{c\in\mathcal{C}} {s_{i,c}[k] \sigma_c^2}}\nonumber\\
\quad\text{subject}&~\text{to}~(\ref{nodel_flow})-(\ref{tank_4}), (\ref{pipe_pump2})-(\ref{pipe_nopump2}), (\ref{new_H_G11})-(\ref{disct4}), (\ref{lin1})-(\ref{disct41}) \nonumber\\
&\sum_{(i,j)\in\mathcal{P}}{\frac{\rho g \delta}{\eta_{i,j}} \sum_{b\in\mathcal{B}} {\Phi_{i,j,b}[k] \zeta_c}}\le r[k]\delta\nonumber
\end{align}
where $\rho[k]=[z[k]^\top,s[k]^\top,\Phi[k]^\top]^\top$. Problem ${\textbf{N}}_{4}$ is a mixed-integer second-order cone program that can be solved efficiently with commercial solvers such as GUROBI. In addition, several efficent algorithms for solving mixed-integer second-order cone programs exist (e.g., see \cite{MISCOP1} and \cite{MISCOP2}).

The solution to ${\textbf{N}}_{4}$ provides an approximation to an exact solution to ${\textbf{N}}_{1}$. The quality of this approximation can be improved by increasing the number of discrete values that each continuous variable, $H_{i,j}[k]$ $(i,j)\in\mathcal{P}$ and $H_i[k]$ for all $i\in\mathcal{T}$, can take. In addition, if ${\textbf{N}}_{2}$ and ${\textbf{N}}_{3}$ are equivalent, the water network operator can use the solution to ${\textbf{N}}_{4}$ to schedule its tanks, pumps, and valves in a centralized fashion. However, the water network operator needs to construct an exact solution to ${\textbf{N}}_{2}$ from a solution to ${\textbf{N}}_{4}$ in order to control the water network optimally. Corollary~\ref{cor2} enables the operator to construct an exact solution to ${\textbf{N}}_{2}$ from an exact solution to ${\textbf{N}}_{4}$ when the conditions of Theorem \ref{thrm3} are satisfied. The proof closely follows the proof of Theorem \ref{thrm3}. Due to space limitations, we do not present the proof.

\begin{cor}\label{cor2}
Given the parameters $\overline{V}_i$, $\underline{H}_i$, $H^0_i$, $\overline{\omega}_{i,j}$, $\eta_{i,j}$, $f^d_{i,j}$, and $D[k]$, let $x^*[k]$, $\rho^*[k]$, and $w^*[k]$ denote the optimal solution to ${\textbf{N}}_{4}$. Consider vectors $x^{(r)}[k]$, $\rho^{(r)}[k]$, and $w^{(r)}[k]$ with the following entries:
\begin{align}
V^{(r)}[k]&=V^*[k],~Q^{(r)}[k]=Q^*[k],\rho^{(r)}[k]=\rho^*[k],\nonumber\\
{H_{i}^{(r)}[k]}&={H_{i}^*[k]}+\psi_{i}[k],~\forall i\in\mathcal{N}_y,\nonumber\\
{W_{i,j}^{(r)}[k]}&={({Q^*_{i,j}[k]})}^2,~\forall(i,j)\in\mathcal{L},\nonumber\\
H_{i,j}^{(r)}[k]&=f^d_{i,j} {({Q^*_{i,j}[k]})}^2,~\forall(i,j)\in\mathcal{L},\nonumber\\
H_{i,j}^{(r)}[k]&=H_{i,j}^*[k],~\forall(i,j)\in\mathcal{P},\nonumber\\
{H}^{(r)}_{i,j}[k]&=\left\{ \begin{array}{ll}
{H_{i,j}^*[k]}+\psi_{i}[k],~\text{if}~j\in\mathcal{M} ~\text{and}~ (i,j)\in\mathcal{V}\nonumber\\
{H_{i,j}^*[k]},~\text{if}~j\notin\mathcal{M}~\text{and}~ (i,j)\in\mathcal{V}\nonumber
\end{array} \right.\nonumber
\end{align}
Then, vectors $x^{(r)}[k]$, $\rho^{(r)}[k]$, and $w^{(r)}[k]$ form an exact solution to ${\textbf{N}}_{4}$.
\end{cor}

Recent demonstrations of PRVs have shown their benefits in control capabilities \cite{valv1}-\!\!\cite{valv4}, and have incentivized water network operators to invest in PRVs. In the near future, we envision PRVs will be available at all junctions with multiple incoming-outgoing pipes. When the network topology does not satisfy the conditions of Theorem \ref{thrm3}, ${\textbf{N}}_{4}$ will not be necessarily exact. For such topologies, we can use a solution to ${\textbf{N}}_{4}$ as an upper bound on the objective function in ${\textbf{N}}_{1}$ to evaluate the performance of different energy harvesting heuristics.


\section{Numerical Results}\label{num_sec}
\begin{figure}[t]
\begin{center}
\includegraphics[width=2.8in]{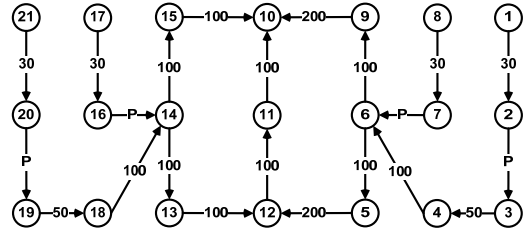} \caption{A water network comprising 21 nodes (circles), multiple pipes (arrows), and pumps (arrows with label P). The length of each pipe (in m) is shown on the arrow.} \label{system_model162}
\end{center}
\end{figure}

Consider a water supply network comprised of 15 junctions, 4 reservoirs, 4 pumps, and 2 tanks, as shown in Fig. \ref{system_model162}. The water network operator uses four variable speed pumps with the parameters $a_{i,j}=-1.0941\times 10^{-4}$, ${b}_{i,j}=5.1516\times10^{-2}$, and ${c}_{i,j}=223.32$ to pump water from the two reservoirs to different nodes in the system \cite{pump_group}. In addition, the operator has installed PRVs in each junction with multiple incoming pipes, and has selected the minimum allowable pressure head at each junction to be equal to 5~m. We assume that the elevation head of each node is zero except nodes 7, 8, 16, and 17 at which we have ${H}_{7}^0={H}^0_{8}={H}^0_{16}={H}^0_{17}=-10$~m. We further assume that the diameter and the friction factor of each pipe $(i,j)$ are equal to $d_{i,j} = 0.3$ (m) and $f^d_{i,j} = 0.001$, respectively. We select the cross-sectional wetted area, the capacity, and the initial volume of water of each tank $i$ to be equal to $A_i=\pi\times(25/2)^2$, $\overline{V}_i = 30 A_i$, and $V_i[0]=0.2\overline{V}_i$, respectively. A similar network topology is studied in \cite{dfooladi1}, \cite{model_numerical}.

We focus on a period of length $T=1$ hour, and assume that the time is divided into slots of size $\delta=5$ minutes (i.e., the number of time slots is $K=12$).
\textcolor{black}{As mentioned earlier, we consider water demands as stochastic parameters that are constant over the duration of a time slot, and assume that at the beginning of each time slot $k$, the water network operator can perfectly estimate the value of water demand vector $D[k]$. We select the values of the water demands randomly as shown in Table \ref{bounds}. Note that $\text{U}[a,b]$ denotes a random variable that is uniformly distributed in the interval $[a,b]$. Here, we assume $\overline{r}=200$ KWatt.}
As mentioned earlier, we assume that water demands are constant over the duration of a time slot. We select the values of the water demands randomly as shown in Table \ref{bounds}. Note that $\text{U}[a,b]$ denotes a random variable that is uniformly distributed in the interval $[a,b]$. Here, we assume $\overline{r}=200$ KWatt.


To demonstrate the effectiveness of our energy harvesting scheme, we consider a sequence of demand response signal values generated randomly from the interval $[0,\overline{r}]$. We use our solution techniques with the design parameters $\zeta_0=\sigma_0=0$, $\zeta_B=40$~m, $\sigma_C=30$~m, $B=80$, and $C=60$ to compute the optimal set-points of the pumps and valves as well as the optimal water flows through the network. Note that, in each junction with multiple incoming pipes, the incoming pipes are equipped with PRVs. Hence, the network topology satisfies the conditions of Theorem \ref{thrm3}. To compute the optimal schedules of pumps, tanks, and valves, we solved ${\textbf{M}}_{1}$ and ${\textbf{N}}_{1}$ using the proposed solution techniques. We were able to compute exact solutions to ${\textbf{M}}_{1}$ and ${\textbf{N}}_{1}$ using GUROBI in few seconds. This demonstrates the effectiveness of our solution techniques against the original problem formulation.

\begin{table}
    \caption{Network nodes: type and demand ($\text{m}^3/\text{hour}$).}
    \vspace{-0.15in}
    \label{bounds}
    \begin{center}
        \resizebox {0.35\textwidth }{!}
        {\begin{tabular}{|c|c|c||c|c|c|}
        \hline
        Node  & Type  & Demand &Node  & Type  & Demand \\
        \hline
        \hline
        1 & reservoir & 0 & 2 & junction& 0 \\
        \hline
        3 & junction&  0& 4 & tank& 0\\
        \hline
        5 & junction&  0& 6 & junction& 0 \\
        \hline
        7 & junction& 0& 8 & reservoir&  0\\
        \hline
        9 & junction& U[576,864]& 10 & junction&  U[720,1008]\\
        \hline
        11 & junction& U[432,720]& 12 & junction& U[864,1080]\\
        \hline
        13 & junction&  0& 14 & junction&  0\\
        \hline
        15 & junction&  U[360,648]& 16 & junction& 0 \\
        \hline
        17 & reservoir& 0& 18 & tank& 0 \\
        \hline
        19 & junction&  0& 20 & junction& 0 \\
        \hline
        21 & reservoir& 0&  &  & \\
        \hline
        \end{tabular}}
    \end{center}
\vspace{-0.1in}
\end{table}



\begin{figure}
\begin{center}
\includegraphics[width=2.8in]{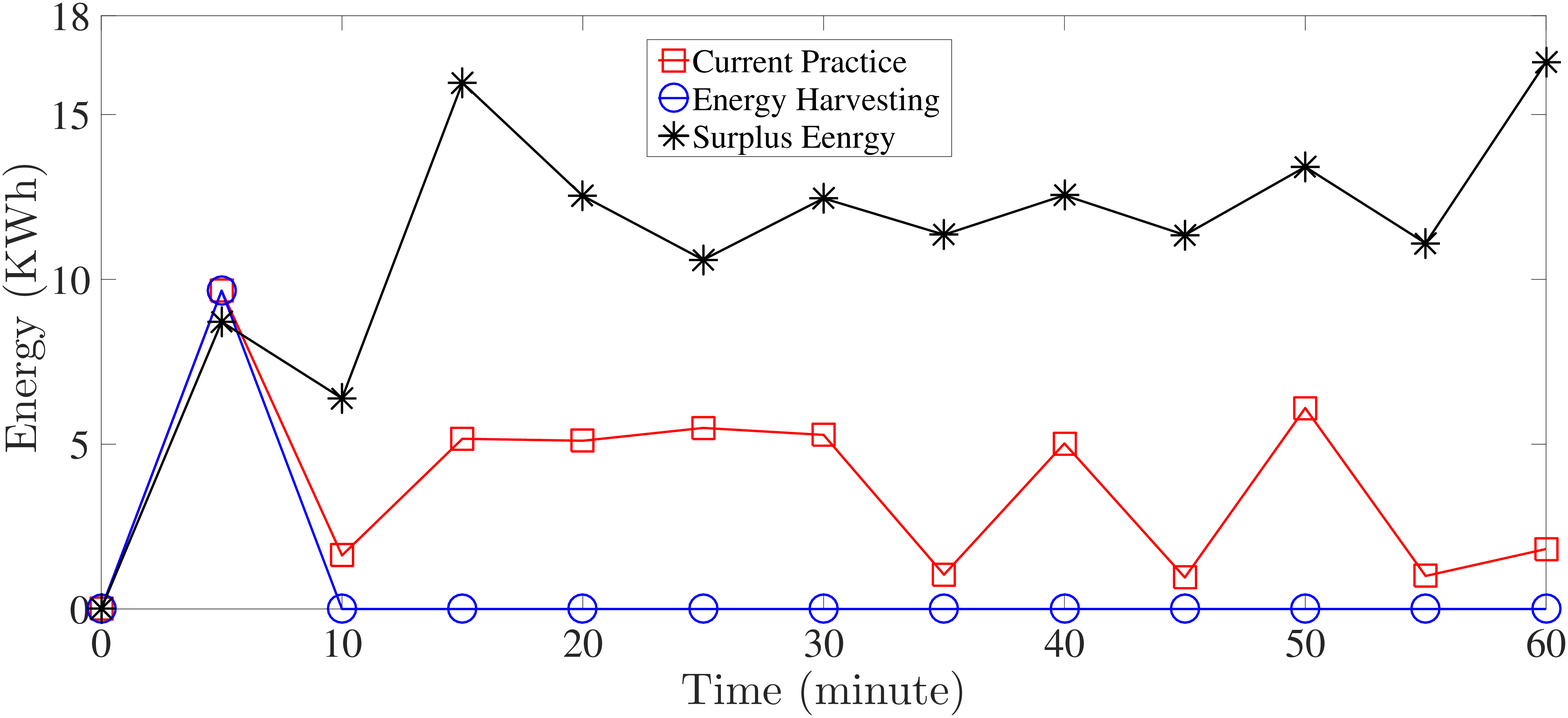}\caption{The energy consumption of the water network as a function of time: The graph in red (dashed line with squares) shows the energy consumption of the water network without any energy harvesting mechanism while the graph in blue (dashed line with circles) shows the energy consumption of the system when the operator is harvesting the surplus energy. The graph in black represents the surplus energy provided by the power system.}\label{tankT1}
\end{center}
\begin{center}
\includegraphics[width=2.8in]{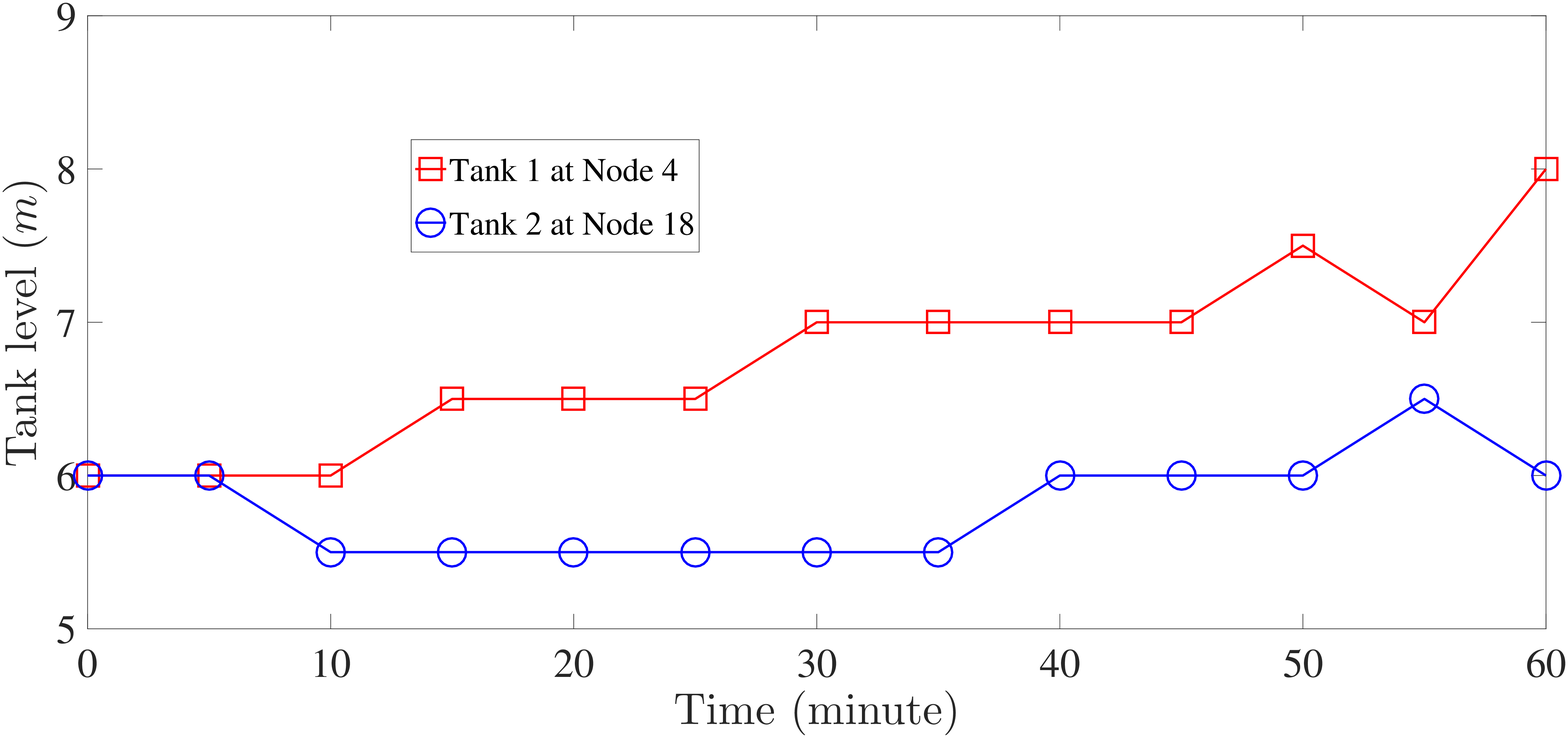}\caption{The height of water in the tanks as a function of time.}\label{tankT}
\end{center}
\end{figure}

Our numerical results are shown in Figs. \ref{tankT1}-\ref{tankT}. The results in Fig. \ref{tankT1} show that the water network operator can significantly reduce the amount of energy that it needs to purchase from the electricity market, by participating demand response programs. The results in Fig. \ref{tankT} show that the water level of the tank at node 4 is increasing in time. We observed the same behavior for the water level of the tank at node 18. \textcolor{black}{Our results show that the water network operator can harvest the surplus energy resulting
from renewable-based generations in the tanks. Note that ISOs typically curtail the surplus renewable energy. Therefore, the proposed framework potentially enables  ISOs to utilize their renewable energy resources more efficiently, and to reduce fossil-fuel based generations. In summary, there are several benefits with the proposed architecture in this paper: (i) the water network operator can significantly minimize its electricity costs, and (ii) the electrical grid operator can potentially use its renewable resources more efficiently.}

\vspace{-0.15cm}

\section{Conclusion}\label{concl_sec}
In this paper, we focused on the problem of energy harvesting in water supply networks, and formulated two joint optimal pump scheduling and water flow problems that are NP-hard. We proposed a convex relaxation and an approximation technique that enable us to transform the proposed problems into mixed-integer second-order cone programs. \textcolor{black}{We provided the conditions under which the proposed relaxation is exact, and provided an algorithm for constructing a solution to the original problem from a solution to the relaxed problem. Finally, using a real-world water supply network, we demonstrated the effectiveness of our relaxation and approximation techniques in solving the energy harvesting optimization problems.}

\vspace{-0.2cm}

\appendix
In this appendix, we provide the proof of Theorem \ref{thrm3}. We begin with introducing some notations and definitions. %

\vspace{-0.2cm}
\subsection{Preliminaries}
\textcolor{black}{Consider graph $\mathcal{G}_y(\mathcal{N}_y,\mathcal{E}_y\cup\mathcal{E}^{(f)}_y)$ where $\mathcal{N}_y\subset\mathcal{N}$, $\mathcal{E}_y\subset\mathcal{E}$, and $\mathcal{E}^{(f)}_y\subset\mathcal{E}^{(f)}$.} Given node $i\in\mathcal{N}_y$, the set of nodes $\mathcal{N}_y$ can be divided into five disjoint subsets as follows:
\[\mathcal{N}_y=\mathcal{U}_{i}\cup\mathcal{D}_{i}\cup\mathcal{U}^{c}_{i}\cup\mathcal{D}^{c}_{i}\cup\{i\},\]
where $\mathcal{U}_{i}$ and $\mathcal{D}_{i}$ denote the sets of upstream and downstream nodes of node $i$, respectively, in graph $\mathcal{G}_y$. $\mathcal{U}^c_{i}$ (resp. $\mathcal{D}^c_{i}$) denotes the set of nodes that are connected to the upstream (resp. downstream) nodes of node $i$, but there does not exist any path between them and node $i$. However, for each node $j_1\in\mathcal{U}^c_{i}$ (resp. $j_1\in\mathcal{D}^c_{i}$), there exist a node $j_2\in\mathcal{U}_{i}$ (resp. $j_2\in\mathcal{D}_{i}$) and a path between nodes $j_1$ and $j_2$. Notice that directed graph $\mathcal{G}_y$ is connected, i.e., there exists a path between nodes $j_1$ and $j_2$ if $j_1\in\mathcal{D}_{j_2}$. We further define the operator $\text{deg}_{\text{I}}(\mathcal{G}_y)$ that returns the maximum indegree over all the vertices of graph $\mathcal{G}_y$.

\vspace{-0.5cm}

\subsection{Proof of Theorem \ref{thrm3}}
We define an auxiliary variable $\nu[k]$ for $k\in\mathcal{K}$, and select $\nu[k]$ to be equal to $\sum_{i\in\mathcal{T}} {A_i} H_{i}^2[k]$, i.e., $\nu[k]$ equals the objective function in ${\textbf{N}}_{3}$.

Let the schedules $\nu^*[k]$, $x^*[k]$, and $w^*[k]$ denote the optimal solution to ${\textbf{N}}_{3}$. It can be verified that the schedules $Q^*[k]$ and $V^*[k]$ form a set of feasible water flows for the water network. However, the set of pressure heads $H^*[k]$ and $G^*[k]$, and $w^*[k]$ are not necessarily feasible for the system since in ${\textbf{N}}_{3}$, constraint (\ref{head_loss}) is relaxed. Our goal is to show that under the conditions of Theorem \ref{thrm3}, we can always construct a new set of schedules $\nu^{(r)}[k]$, $x^{(r)}[k]$, and $w^{(r)}[k]$ that is feasible to ${\textbf{N}}_{3}$ with $\nu^{(r)}[k]=\nu^*[k]$. To do so, let us assume that there exists a pipe $(i',j')\in\mathcal{L}$ for which ${{({Q_{i',j'}^*[k]})}^2} < {W^*_{i',j'}[k]}$ with $Q_{i',j'}^*[k]\neq0$, i.e., assume that ${\textbf{N}}_{3}$ is not exact. Since \textcolor{black}{$\mathcal{G}_y(\mathcal{N}_y,\mathcal{E}_y\cup\mathcal{E}^{(f)}_y)$ is loop-free}, two cases can be considered here:

\textbf{Case 1:} Let us assume that $\text{deg}_{\text{I}}(\mathcal{G}_y)=1$. Therefore, pipe $(i',j')$ is not in a parallel path between two nodes. Using $H^*[k]$, $G^*[k]$, and $w^*[k]$, we can construct a new set of pressure heads $H_i^{(r)}[k]$ and $H_{i,j}^{(r)}[k]$, and $W_{i,j}^{(r)}[k]$ as follows:
\begin{align}\small
&{H}^{(r)}_{i}[k]=\left\{ \begin{array}{ll}
{H^*_{i'}[k]},~\text{if}~i=i'\nonumber\\
{H_{i}^*[k]}+\epsilon_{i',j'},~\text{if}~i\in\mathcal{D}_{i'}~\text{and}~\mathcal{N}_{j',i}=\emptyset\nonumber\\
{H_{i}^*[k]},~\text{if}~i\in\mathcal{D}_{i'}~\text{and}~\mathcal{N}_{j',i}\neq\emptyset\nonumber\\
{H_{i}^*[k]},~\text{if}~i\in\mathcal{U}_{i'}\cup\mathcal{U}^c_{i'}\nonumber
\end{array} \right.\\
&{W}^{(r)}_{i,j}[k]=\left\{ \begin{array}{ll}
{({Q^*_{i,j}[k]})}^2,~\text{if}~i=i'~\text{and}~j=j'\nonumber\\
{W}^*_{i,j}[k],~\text{otherwise}\nonumber
\end{array} \right.,\forall(i,j)\in\mathcal{L},\nonumber\\
&H_{i,j}^{(r)}[k]=f^d_{i,j} {{W^{(r)}_{i,j}[k]}},~\forall(i,j)\in\mathcal{L},\nonumber\\
&H_{i,j}^{(r)}[k]=H_{i,j}^*[k],~\forall(i,j)\in\mathcal{P}\cup\mathcal{V},\nonumber
\end{align}
where $\epsilon_{i',j'}=f^d_{i',j'} ({W_{i',j'}^*[k]}-{({Q^*_{i',j'}[k]})}^2)>0$. Recall that $\mathcal{N}_{i,j}$ is the set of tanks, and junctions with multiple incoming pipes over all the paths between nodes $i$ and $j$ including $j$. Note that $\mathcal{D}^c_{i'}=\emptyset$ and there exists at most one path between nodes $i$ and $j$ since $\text{deg}_{\text{I}}(\mathcal{G}_y)=1$. It can be verified that the schedules $Q^*[k]$, $V^*[k]$, $H^{(r)}[k]$, $G^{(r)}[k]$, and $w^{(r)}[k]$ form a feasible solution to ${\textbf{N}}_{3}$.

\textbf{Case 2:} Let us assume that $\text{deg}_{\text{I}}(\mathcal{G}_y)>1$, and that pipe $(i',j')$ is not in a parallel path between two nodes. Using $H^*[k]$, $G^*[k]$, and $w^*[k]$, we can construct a new set of pressure heads $H_i^{(r)}[k]$ and $H_{i,j}^{(r)}[k]$, and $W_{i,j}^{(r)}[k]$ as follows:
\begin{align}\small
&{H}^{(r)}_{i}[k]=\left\{ \begin{array}{ll}
{H^*_{i'}[k]},~\text{if}~i=i'\nonumber\\
{H_{i}^*[k]}+\epsilon_{i',j'},~\text{if}~i\in\mathcal{D}_{i'}~\text{and}~\mathcal{N}_{j',i}=\emptyset\nonumber\\
{H_{i}^*[k]},~\text{if}~i\in\mathcal{D}_{i'}~\text{and}~\mathcal{N}_{j',i}\neq\emptyset\nonumber\\
{H_{i}^*[k]},~\text{if}~i\in\mathcal{U}_{i'}\cup\mathcal{U}^c_{i'}\cup\mathcal{D}^c_{i'}\nonumber
\end{array} \right.\\
&{W}^{(r)}_{i,j}[k]=\left\{ \begin{array}{ll}
{({Q^*_{i,j}[k]})}^2,~\text{if}~i=i'~\text{and}~j=j'\nonumber\\
{W}^*_{i,j}[k],~\text{otherwise}\nonumber
\end{array} \right.,\forall(i,j)\in\mathcal{L},\nonumber\\
&H_{i,j}^{(r)}[k]=f^d_{i,j} {{W^{(r)}_{i,j}[k]}},~\forall(i,j)\in\mathcal{L},\nonumber\\
&H_{i,j}^{(r)}[k]=H_{i,j}^*[k],~\forall(i,j)\in\mathcal{P},\nonumber
\end{align}
where $\epsilon_{i',j'}=f^d_{i',j'} ({W_{i',j'}^*[k]}-{({Q^*_{i',j'}[k]})}^2)>0$. For each $(i,j)\in\mathcal{V}$, ${H}^{(r)}_{i,j}[k]$ is given by
\begin{align}\small
&{H}^{(r)}_{i,j}[k]=\left\{ \begin{array}{ll}
{H_{i,j}^*[k]}+\epsilon_{i',j'},~\text{if}~j\in\mathcal{D}_{i'}\cap\mathcal{M}~\text{and}~\mathcal{N}_{j',i}=\emptyset\nonumber\\
{H_{i,j}^*[k]},~\text{otherwise}\nonumber
\end{array} \right.\nonumber
\end{align}
Recall that $\mathcal{N}_{i,j}$ is the set of tanks, and junctions with multiple incoming pipes over all the paths between nodes $i$ and $j$ including $j$, and $\mathcal{M}\subset\mathcal{J}$ is the set of junctions with multiple incoming pipes in \textcolor{black}{graph $\mathcal{G}_y(\mathcal{N}_y,\mathcal{E}_y\cup\mathcal{E}^{(f)}_y)$}. Notice that we can select ${H}^{(r)}_{i}[k]={H_{i}^*}[k]$ for all $i\in\mathcal{D}^c_{i'}$ with $\mathcal{N}_{j',i}\neq\emptyset$ since in each junction $j$ with multiple incoming pipes, the incoming pipes are equipped with PRVs. It can be verified that the schedules $Q^*[k]$, $V^*[k]$, $H^{(r)}[k]$, $G^{(r)}[k]$, and $w^{(r)}[k]$ form a feasible solution to ${\textbf{N}}_{3}$.

In \textcolor{black}{graph $\mathcal{G}_y(\mathcal{N}_y,\mathcal{E}_y\cup\mathcal{E}^{(f)}_y)$}, we may have multiple parallel paths between two arbitrary nodes $i,j\in\mathcal{N}_y$ since $\text{deg}_{\text{I}}(\mathcal{G}_y)>1$. Let us assume that pipe $(i',j')$ is in a path (we call it $\mathcal{P}_1$) from node $i$ to node $j$, and that this path is parallel to another path (we call it $\mathcal{P}_2$) from $i$ to $j$. To ensure that the new schedule $x^{(r)}[k]$, and $w^{(r)}[k]$ is a feasible solution to ${\textbf{N}}_{3}$, we only need to show that this schedule preserves the energy conservation constraint between the parallel paths $\mathcal{P}_1$ and $\mathcal{P}_2$. Let $j''$ denote the first downstream node of node $j'$ with multiple incoming pipes along path $\mathcal{P}_1$, and let $(i'',j'')$ denote the PRV connected to node $j''$. Such a node exists among the downstream nodes of node $j'$ in $\mathcal{P}_1$ since there are two parallel paths from $i$ to $j$. We select ${H}^{(r)}_{i'',j''}[k]={H_{i'',j''}^*[k]}+\epsilon_{i',j'}$ and ${H}^{(r)}_{j''}[k]={H}^*_{j''}[k]$ which imply that each downstream node of node $j''$ will maintain the same pressure head as the one in $x^*[k]$ and $w^*[k]$. Therefore, the new schedule $x^{(r)}[k]$ and $w^{(r)}[k]$ preserves the energy conservation constraint between the parallel paths $\mathcal{P}_1$ and $\mathcal{P}_2$, and hence it is a feasible solution to ${\textbf{N}}_{3}$.

In each of the cases above, we constructed a new feasible solution to ${\textbf{N}}_{3}$. In this solution, we have $Q^{(r)}_{i,j}[k]=Q^*_{i,j}[k]$ for all $(i,j)\in\mathcal{E}$, and hence we obtain $H_i^{(r)}[k]=H_i^*[k]$ for all $i\in\mathcal{T}$. Since the objective function in ${\textbf{N}}_{3}$ is only a function of pressure heads $H_i[k]$'s where $i\in\mathcal{T}$, we have a feasible solution to ${\textbf{N}}_{3}$ with $\nu^{(r)}[k]=\nu^*[k]$. Hence, the schedule $x^{(r)}[k]$ and $w^{(r)}[k]$ is an optimal solution to ${\textbf{N}}_{3}$. Now, if there still exists another pipe $(i',j')\in\mathcal{L}$ (in the new schedule $x^{(r)}[k]$ and $w^{(r)}[k]$) for which ${{({Q_{i',j'}^*[k]})}^2} < {W^{(r)}_{i',j'}[k]}$ with $Q_{i',j'}^*[k]\neq0$, we can follow the same procedure to obtain a feasible solution that is exact for ${\textbf{N}}_{3}$. Therefore, for water networks that satisfy one of the conditions of Theorem \ref{thrm3}, either the computed solution is exact or a new feasible solution to ${\textbf{N}}_{3}$ with the same optimal objective value can be constructed by using the computed solution. This completes the proof.\qed

\end{document}